\documentclass[12pt]{iopart}

\usepackage[colorlinks=true,allcolors=blue]{hyperref}  
\usepackage[hypcap]{caption} 
\usepackage{graphicx,braket}

\begin{document}

\title{Optimization of a solid-state electron spin qubit using Gate Set Tomography}

\author{Juan P. Dehollain$^{1}$\footnote{Present address: QuTech \& Kavli Institute of Nanoscience, TU Delft, 2628 CJ Delft, The Netherlands}, Juha T. Muhonen$^{1}$\footnote{Present address: Center for Nanophotonics, FOM Institute AMOLF, 1098 XG, Amsterdam, The Netherlands}, Robin Blume-Kohout$^{2,3}$, Kenneth M. Rudinger$^{2,3}$, John King Gamble$^{2,3}$, Erik Nielsen$^{2}$, Arne Laucht$^{1}$, Stephanie Simmons$^{1}$\footnote{Present address: Department of Physics, Simon Fraser University, Burnaby, BC, Canada V5A 1S6}, Rachpon Kalra$^{1}$, Andrew S. Dzurak$^{1}$ and Andrea Morello$^{1}$}
\address{$^1$ Centre for Quantum Computation and Communication Technologies, School of Electrical Engineering and Telecommunications, UNSW Australia, Sydney, New South Wales 2052, Australia}
\address{$^2$ Sandia National Laboratories, Albuquerque, New Mexico 87185, USA}
\address{$^3$ Center for Computing Research, Sandia National Laboratories, Albuquerque, New Mexico 87185, USA}
\eads{\mailto{j.p.dehollainlorenzana@tudelft.nl}, \mailto{enielse@sandia.gov}, \mailto{a.morello@unsw.edu.au}}

\begin{abstract}
State of the art qubit systems are reaching the gate fidelities required for scalable quantum computation architectures. Further improvements in the fidelity of quantum gates demands characterization and benchmarking protocols that are efficient, reliable and extremely accurate. Ideally, a benchmarking protocol should also provide information on how to rectify residual errors. Gate Set Tomography (GST) is one such protocol designed to give detailed characterization of as-built qubits. We implemented GST on a high-fidelity electron-spin qubit confined by a single $^{31}$P atom in $^{28}$Si. The results reveal systematic errors that a randomized benchmarking analysis could measure but not identify, whereas GST indicated the need for improved calibration of the length of the control pulses. After introducing this modification, we measured a new benchmark average gate fidelity of $99.942(8)\%$, an improvement on the previous value of $99.90(2)\%$. Furthermore, GST revealed high levels of non-Markovian noise in the system, which will need to be understood and addressed when the qubit is used within a fault-tolerant quantum computation scheme.
\end{abstract}


\vspace{2pc}

\section{Introduction}
One of the main challenges in the physical implementation of a universal quantum computer lies in designing quantum bits that meet the exquisite operation accuracies demanded by fault-tolerant quantum codes. Sophisticated quantum error correction strategies~\cite{Shor1995,Steane1996,Raussendorf2012} have driven required qubit tolerances down into the realm of experimental possibility; numerical evidence suggests that gate fidelities as low as $99\%$ might be sufficient for fault-tolerant operation~\cite{Knill2005,Fowler2012}. Gate fidelities above this value have already been claimed by several qubit systems, including liquid-state NMR~\cite{Ryan2009}, atomic ions~\cite{Olmschenk2010,Brown2011,Harty2014}, superconducting qubits~\cite{Barends2014} and single spins in semiconductors~\cite{Dolde2014,Veldhorst2014,Muhonen2015}. However, all of these demonstrations have been achieved in single or few-qubit systems and it is likely that further optimization will be required in order to maintain the high fidelities above the fault tolerance threshold as the systems scale up. While problems with low-fidelity qubits can be discerned and addressed easily, improving high-fidelity qubits is more challenging since one must characterize the qubit operation to an ever-increasing degree of accuracy. Quantum Process Tomography (QPT)~\cite{Chuang1997} has been a primary method for characterizing qubit gates. By preparing a set of input states, applying the gate to be evaluated to each state and measuring the output states via quantum state tomography, the operator ($G$) corresponding to the applied gate can be extracted. The problem with this method is that it assumes perfect state preparation and measurement (SPAM); therefore, the accuracy in $G$ is limited by the ratio of SPAM to gate errors~\cite{Chow2009,Kim2015}. Most common quantum error correction codes require much higher fidelity on the qubit logic gates than on SPAM~\cite{Knill2005,Fowler2012}. The experimental push to increase gate fidelities without the need to improve as much in SPAM, is rendering QPT obsolete as a means to characterize qubit gates. Randomized benchmarking (RB)~\cite{Emerson2005,Knill2008} is an alternative protocol for assessing the performance of qubit gates. Random gate sequences are applied to the qubit and the measurement outcome is compared to the expected result to obtain an average gate fidelity. By observing the survival probability as the number of gates in the sequences are increased, we can extract an average gate fidelity which is independent of SPAM. The downside to this protocol is that it outputs a single benchmark for qubit gate performance, without providing further insight into qubit characteristics and the nature of the errors.

Gate Set Tomography (GST)~\cite{Blume2013} is a tool for characterizing logic operations in a qubit system. By analysing carefully constructed experiments consisting of state preparation, quantum operation sequences, and  measurements, it self-consistently characterizes the experimental system. GST operates with minimal assumptions about physical characteristics of the system; it outputs a set of logical gate operators---a gate set---that models the behaviour of the device. Characteristics of the system relevant to quantum information processing can be directly extracted from the gate set, such as rotation angles, relaxation and dephasing rates, and randomized benchmarking decay rates. By computing the goodness of a GST fit (i.e. how well the model fits the experimental data), one reveals any deviation in the behaviour of the device from an ideal qubit system. GST has been previously implemented in a solid-state charge qubit~\cite{Kim2015}, as a means to extract the process fidelity of the qubit gates.

Here we make use of the high-accuracy gate characterization provided by GST, in order to optimize the performance of a solid-state spin qubit. We first describe the physical system and the experimental methods used to perform a GST analysis of the gate fidelities. Analysing the information extracted by the GST protocol provides us with an opportunity to further optimize the qubit operation. We then complement the GST study with a new RB measurement, which highlights the improved gate fidelity obtained by applying the GST diagnostics. Finally, we discuss the current limitations to the accuracy and reliability of GST and propose future work to address these limitations.

\section{Qubit description and operation}

GST is \emph{architecture-agnostic}, in that it directly characterizes the experimental system in the language of quantum information processing. Hence, to effectively interpret the GST results to help improve the experiment, it is necessary to understand the underlying physics, which we detail below.

\textit{The physical implementation of the qubit logic states} -- The qubit used in this study is the quantum two-level system formed by the spin-$\frac{1}{2}$ states of an electron bound to a $^{31}$P donor, implanted~\cite{Donkelaar2015} in isotopically purified $^{28}$Si~\cite{Itoh2014}. The fabrication and operation of the device has been described in great detail in references~\cite{Morello2009,Morello2010,Pla2012,Pla2013,Muhonen2014}. The spin energy states are split by an externally applied magnetic field $B_0 = 1.55$~T. The electron spin is coupled to the $^{31}$P spin-$\frac{1}{2}$ nucleus via the hyperfine interaction $A = 98$~MHz, resulting in a two-spin, four-level system, whose eigenstates are the product states of the electron and nuclear spins. The relaxation rate of the nuclear spin is orders of magnitude smaller than the electron relaxation rate, allowing us to operate on a two-level electron spin subsystem with the nuclear spin `frozen' in an energy eigenstate. The qubit logic states $\ket{1}$ and $\ket{0}$ are then the eigenstates of the electron spin $\ket{\uparrow}$ and $\ket{\downarrow}$ respectively.

\begin{figure}
 \includegraphics[width=\textwidth]{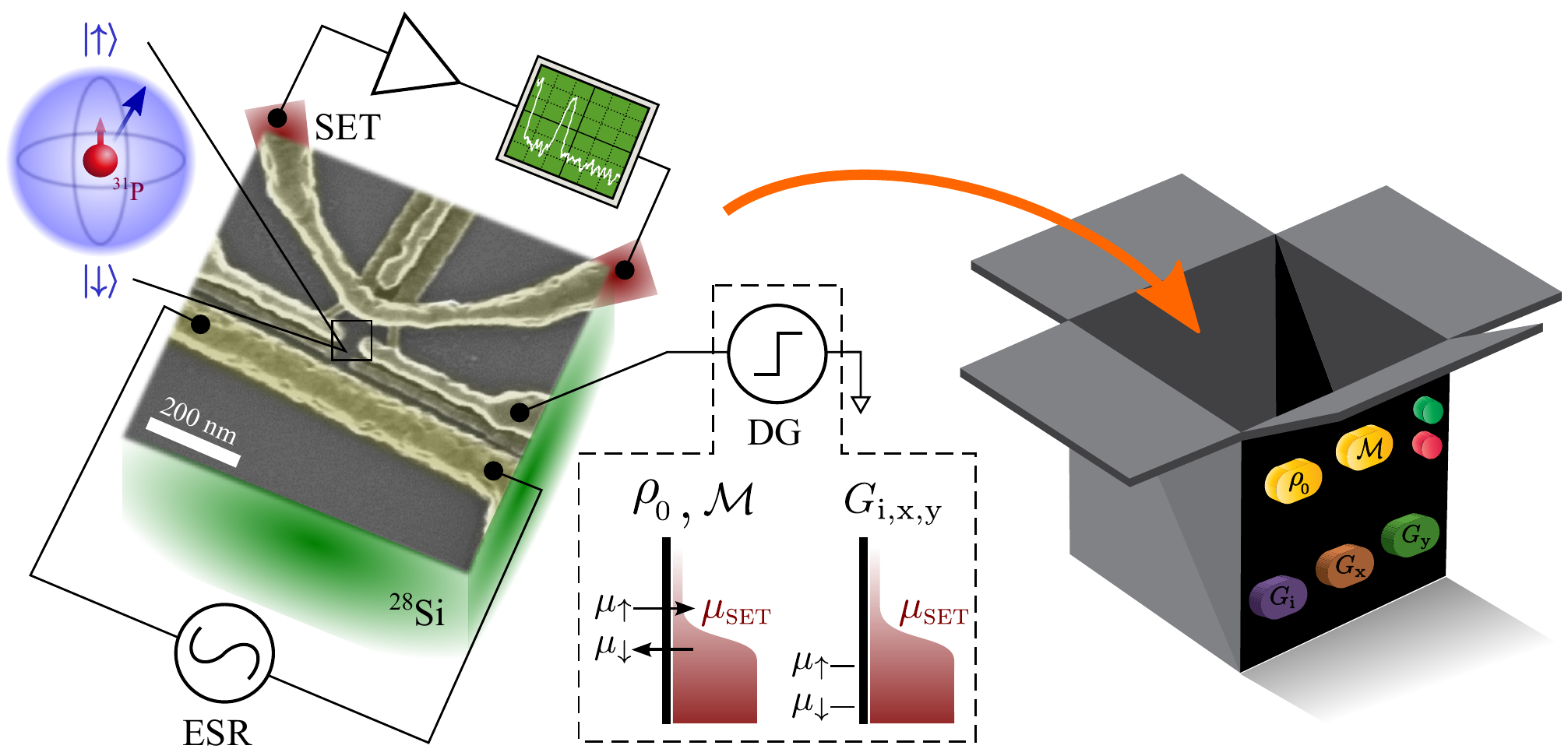}
 \caption{Diagram of qubit device and GST model of a qubit. SEM image of the on-chip gate structure of a device identical to the one used here. The aluminium gates have been false coloured for clarity. Depicted in red are the source-drain n$^+$ regions which connect the SET to the current measurement electronics. For initialization and measurement, the donor gates are pulsed such that $\mu_{\uparrow} > \mu_\mathrm{SET} > \mu_{\downarrow}$, inducing spin-dependent tunnelling between the donor and SET. When applying a gate sequence, the DG are pulsed to higher voltage to prevent the donor electron from tunnelling to the SET. The inset diagram---zoomed from the approximate donor location---represents the Bloch sphere of the qubit, consisting on the spin of an electron confined by an implanted $^{31}$P donor, with its nuclear spin frozen in an eigenstate. The GST model treats the qubit as a black box with buttons which allow to initialize ($\rho_0$), apply each gate in the gate set ($G_\mathrm{i,x,y}$) and measure ($\mathcal{M}$) in the observable basis ($\ket{\uparrow}$ or $\ket{\downarrow}$).\label{fig:device}}
\end{figure}

\textit{State preparation and measurement} are performed via spin dependent tunnelling of the $^{31}$P bound electron to and from a nearby single electron transistor (SET)~\cite{Morello2009,Morello2010}. For this purpose, an aluminium gate stack is fabricated on top of an $8$~nm SiO$_2$ layer, on the surface of the substrate above the donor. The substrate consists of a $1$~$\mu$m epilayer of isotopically purified $^{28}$Si with $800$~ppm residual $^{29}$Si concentration, grown on a natural silicon wafer~\cite{Itoh2014}. The SET accumulates electrons from n$^+$ source-drain regions defined by phosphorus diffusion. The full device structure---as seen in \fref{fig:device}---contains the SET, a set of gates (DG) used to control the electrochemical potential of the donor and an ESR antenna used for qubit state manipulation~\cite{Dehollain2013}. The SET is very sensitive to changes in the electrostatic environment, providing high-fidelity detection of the charge state of the $^{31}$P donor. Its electron island also acts as a reservoir to which the donor is tunnel coupled. The device is cooled down in a dilution refrigerator to an electron temperature $T_\mathrm{e} \approx 100$~mK. At this temperature, the thermal broadening of the Fermi sea in the SET island ($\Delta E_\mathrm{F}$) is much smaller than the Zeeman splitting ($E_\mathrm{Z}$) of the donor spin states. By tuning the donor spin electrochemical potentials ($\mu_{\uparrow,\downarrow}$) with respect to that of the SET island ($\mu_\mathrm{SET}$), such that $\mu_{\uparrow} > \mu_\mathrm{SET} > \mu_{\downarrow}$, we restrict donor$\rightarrow$island tunnelling to a spin-up electron, and island$\rightarrow$donor tunnelling to spin-down electrons~\cite{Morello2010}. This allows us to perform single-shot readout and initialization with fidelities $>98\%$.

\textit{The gate set} -- Logic gates are applied with electron spin resonance (ESR) pulses. An oscillating magnetic field with amplitude $B_1$ and frequency $\nu$, matching the qubit ESR frequency $\nu_0 = \gamma_\mathrm{e} B_0 + A/2 \approx 43$~GHz (where $\gamma_\mathrm{e} = 28$~GHz/T is the electron gyromagnetic ratio), will cause the spin qubit state to rotate coherently between $\ket{\uparrow}$ and $\ket{\downarrow}$. The frequency of rotation $\nu_1$ can be extracted from the Rabi formula as
\begin{eqnarray}
\nu_1 = \sqrt{\left( \nu_0 - \nu \right)^2 + \left( \frac{B_1}{2} \gamma_e \right)^2} \label{eq:omega}
\end{eqnarray}
The x axis in the rotating frame of the qubit is defined by the phase of the first microwave pulse applied to it. Subsequent pulses can be phase-shifted by an angle $\varphi_\mathrm{p}$ to achieve rotations about an axis rotated by $\varphi_\mathrm{p}$ with respect to x.
By controlling $B_1$, the pulse duration $\tau_\mathrm{p}$ and $\varphi_\mathrm{p}$, we can encode any arbitrary qubit state. The device contains an on-chip broadband (DC-$50$~GHz) antenna~\cite{Dehollain2013} used to transmit ESR pulses to the qubit. The antenna is connected to an Agilent~E8267D vector signal generator. The $\sim43$~GHz microwave signal is modulated by its internal dual arbitrary waveform generator, which allows precise and simultaneous control of $B_1$, $\tau_\mathrm{p}$ and $\varphi_\mathrm{p}$. For the experiments presented here, we use a fixed $B_1 \approx 12$~$\mu$T and calibrate $\tau_\mathrm{p}$ and $\varphi_\mathrm{p}$ to apply the desired gate. For the purpose of GST we will characterize two active gates: $G_\mathrm{x}$ and $G_\mathrm{y}$. $G_\mathrm{x}$ corresponds to a $\pi/2$ rotation on the x-axis of the Bloch sphere and is implemented by a pulse with $\tau_{\pi/2} = (4\nu_1)^{-1}$. $G_\mathrm{y}$ is a $\pi/2$ rotation on the y-axis of the Bloch sphere and is implemented by an identical pulse as $G_\mathrm{x}$, but with a relative $\varphi_\mathrm{p} = \pi/2$. Taken together these two gates are informationally complete, since they generate the single-qubit Clifford group. In addition to the active gates, we include the identity gate $G_\mathrm{i}$, where no pulse is applied for the same duration $\tau_{\pi/2}$. This gate characterizes the behaviour of a qubit while it sits idle, waiting for other operations to finish in the quantum processor. The superoperators corresponding to each of these gates are displayed in \tref{tab:targets}.

\begin{table}
\caption{\label{tab:targets}Target superoperators for the experimental gate set in the Pauli basis, with ordering $\mathrm{i}$, $\mathrm{z}$, $\mathrm{x}$, $\mathrm{y}$.}
\begin{indented}
\item[]\begin{tabular}{@{}ll|ll|ll}
\br
$\displaystyle G_\mathrm{i}$&$\displaystyle \left(\begin{array}{cccc}1&0&0&0\\0&1&0&0\\0&0&1&0\\0&0&0&1\end{array}\right)$&
$\displaystyle G_\mathrm{x}$&$\displaystyle \left(\begin{array}{cccc}1&0&0&0\\0&1&0&0\\0&0&0&-1\\0&0&1&0\end{array}\right)$&
$\displaystyle G_\mathrm{y}$&$\displaystyle \left(\begin{array}{cccc}1&0&0&0\\0&0&0&1\\0&0&1&0\\0&-1&0&0\end{array}\right)$\\
\br
\end{tabular}
\end{indented}
\end{table}

\textit{The decoherence rates} -- For the electron spin qubit, the free induction decay and Hahn echo decay times have been measured to be $T_2^* = 0.16$~ms and $T_2 = 1$~ms respectively~\cite{Muhonen2014}. Under constant driving, the qubit can maintain its coherence for up to $T_{1\rho} = 1.3$~s~\cite{Laucht2016}. All of these dephasing times are shorter than the measured spin-lattice relaxation time $T_1 \approx 3$~s.

\section{Gate Set Tomography}
GST~\cite{Blume2013} is a method for characterizing a set of quantum processes (gates), state preparation, and measurement simultaneously. GST requires no pre-calibration, and as such stands in contrast to state tomography, which requires pre-calibrated gates, and process tomography, which requires pre-calibrated state preparation and measurement. Furthermore, GST is able to obtain high-accuracy estimates efficiently, meaning that the number of experiments required to obtain a given accuracy, scales optimally with the desired accuracy. To use GST, one must perform a pre-determined set of experiments. Each experiment consists of 1) state preparation, 2) a sequence of gates, performed one after another, and 3) a measurement. Each gate sequence consists of three parts: 1) a short `fiducial' gate sequence, followed by 2) a `germ' sequence repeated some number of times, followed by 3) another short `fiducial' sequence. Given a set of fiducial sequences, a set of germ sequences, and a list of maximum lengths (which dictate the number of times each germ is repeated), the set of all combinations of (\emph{preparation fiducial}, \emph{germ repeated to max-length}, \emph{measurement fiducial}) gives the complete list of gate sequences required to run GST. Experiments for each gate sequence are repeated multiple times, and the resulting counts of measurement outcomes serve as input to the GST estimation algorithms. These algorithms find the best-fit \emph{gate set} to the experimental data. Because the gate set is defined to contain only single-qubit operations, i.e. operations acting on a 2-dimensional Hilbert state space, a gate set cannot capture effects due to additional Hilbert space dimensions. In particular, memory effects due to the environment, which are an example of what we refer to as `non-Markovian noise', cannot be fit by \emph{any} as-defined gate set. All physical systems will suffer from some degree of non-Markovian noise, and GST can detect this by assessing how well the best-fit gate set is able to reproduce the experimental data. The Pearson chi-squared test and the likelihood-ratio test are used to quantify the `goodness-of-fit'.

The fiducial gate sequences and germ gate sequences, which are used to construct the final list of experiments as explained above, depend upon the ideal desired gates. In our case these gates, given in Table 1, result in the six fiducial sequences
\begin{eqnarray*}
\{(\mathrm{empty}), \enskip G_\mathrm{x}, \enskip G_\mathrm{y}, \enskip G_\mathrm{x}G_\mathrm{x}, \enskip G_\mathrm{x}G_\mathrm{x}G_\mathrm{x}, \enskip G_\mathrm{y}G_\mathrm{y}G_\mathrm{y}\}
\end{eqnarray*}
and eleven germ sequences
\begin{eqnarray*}
\{G_\mathrm{x}, \enskip G_\mathrm{y}, \enskip G_\mathrm{i}, \enskip G_\mathrm{x}G_\mathrm{y}, \enskip G_\mathrm{x}G_\mathrm{y}G_\mathrm{i}, \enskip G_\mathrm{x}G_\mathrm{i}G_\mathrm{y}, \enskip G_\mathrm{x}G_\mathrm{i}G_\mathrm{i},\\
G_\mathrm{y}G_\mathrm{i}G_\mathrm{i}, \enskip G_\mathrm{x}G_\mathrm{x}G_\mathrm{i}G_\mathrm{y}, \enskip G_\mathrm{x}G_\mathrm{y}G_\mathrm{y}G_\mathrm{i}, \enskip G_\mathrm{x}G_\mathrm{x}G_\mathrm{y}G_\mathrm{x}G_\mathrm{y}G_\mathrm{y}\}
\end{eqnarray*}
Details of how fiducial and germ sequences are computed can be found in the supplementary material of reference~\cite{Blume2016}. We used maximum lengths that were increasing powers of two from 1 to 256, which are chosen to include the longest sequences practical on our particular hardware given signal-to-noise and qubit decoherence considerations. The GST analysis was performed using the open-source pyGSTi code~\cite{pyGSTi}.

\section{Optimizing the qubit operation with GST}

\begin{figure}
 \includegraphics[width=\textwidth]{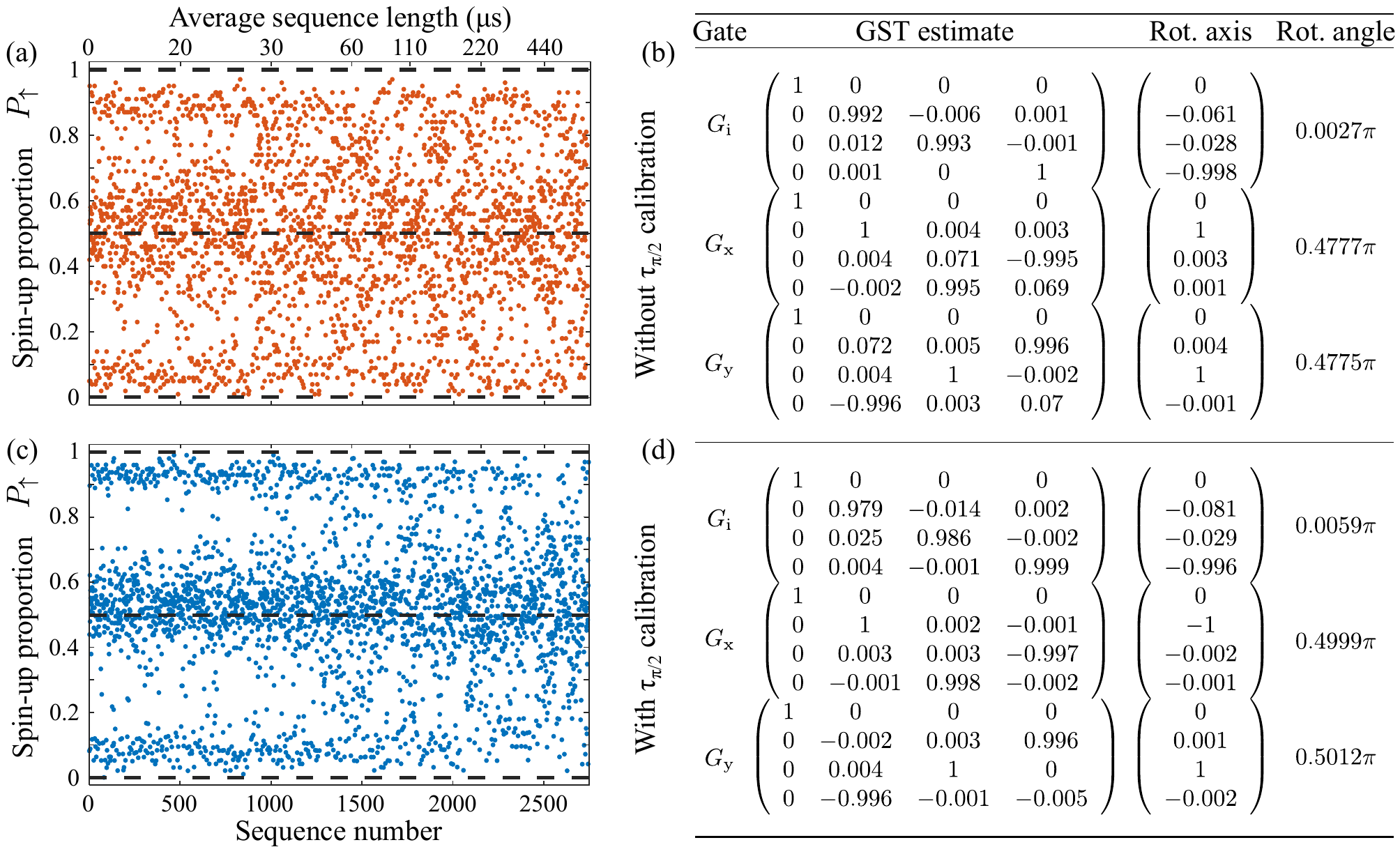}
 \caption{GST results. (a) Raw data points obtained after implementing each of the designed gate sequences and repeating them $100$ times to extract the spin-up proportion $P_\uparrow$ for each sequence. We number the sequences from $0$ to $2736$ as shown in the bottom axis labels, and they increase in length as shown in the top axis labels. Dashed lines show target outcomes for an ideal qubit. (b) Post-processed GST results including the gate operators extracted from the data, and the rotation axis and angle implied by these operators. (c,d) GST data and results after optimizing the pulse length calibration protocol to improve the $\tau_{\pi/2}$ accuracy.\label{fig:gst_results}}
\end{figure}

Each cycle of initialization, gate sequence and measurement was repeated $100$~times for each of the $2737$ sequences constructed for GST. The number of $\ket{\uparrow}$ measurement outcomes was recorded for each sequence and the results were fed back to pyGSTi for analysis. \Fref{fig:gst_results}(a) shows a plot of the spin-up fraction $P_\uparrow$ for all the pulse sequences applied. For an ideal qubit, a sequence can have one of three possible $P_\uparrow$ outcomes: $0$, $0.5$, $1$ (since the gates in our gate set consist of $\pi/2$ rotations). The high-precision of the GST protocol is obtained by designing sequences that amplify gate errors. This error amplification is evident from the scatter around the three $P_\uparrow$ values in the experimental dataset. \Fref{fig:gst_results}(b) shows a table with the estimated gates extracted from GST, highlighting on a separate column the rotation angle implicit in these gate operators. Both $G_\mathrm{x}$ and $G_\mathrm{y}$ show rotation angles of $0.478\pi$, which corresponds to a $4.4$\% under-rotation from the optimal $0.5\pi$. Prior to the development of GST, we performed a qubit optimization using the randomized benchmarking protocol~\cite{Muhonen2015}. RB returns a value for gate fidelity but does not provide any characterization of the gates. Therefore, qubit optimization is achieved by performing sweeps of intuitively chosen qubit operation parameters and searching for the parameter combination which yields the highest gate fidelity. In the RB study, we analysed gate fidelities for different pulse shapes, ESR signal amplitudes and rise times of the pulses. We found a maximum Clifford gate fidelity $\mathcal{F}_\mathrm{G} = 99.90(2)$\% for square pulses, with a rise time of $100$~ns and $B_1 = 12$~$\mu$T (corresponding to $\tau_\pi = 3$~$\mu$s). However, in that study we did not correctly account for the fact that the fixed rise times imply that the area under the time-dependent pulse amplitude---which determines the rotation---is not linear with pulse length. This effect is insignificant for long pulse lengths, but becomes more noticeable as $\tau_\mathrm{p}$ becomes comparable to the rise time. This calibration protocol was designed to only calibrate $\tau_\pi$ and, for the rise time and pulse lengths used in our experiment, $\tau_\pi/2$ is $4.4$\% shorter in rotation than $\tau_{\pi/2}$, as identified by GST.

We corrected the issue by including a separate $\tau_{\pi/2}$ calibration step in the protocol. The data plot in \fref{fig:gst_results}(c)---taken after implementing the optimized calibration protocol---shows significantly less scatter in the data, a first indication that the gates are closer to the target gates. This is confirmed by the GST results in \fref{fig:gst_results}(d), now indicating $G_\mathrm{x}$ and $G_\mathrm{y}$ rotations within $0.7$\% of the target.

The ancillary files contain the full GST reports generated by pyGSTi. Additionally, we have supplied the data files constructed from the experiments, along with the Python notebook used to generate the report. Instructions on how to use these files to generate the reports can be found in the pyGSTi project website~\cite{pyGSTi}.

\begin{figure}
 \centering
 \includegraphics[scale=0.8]{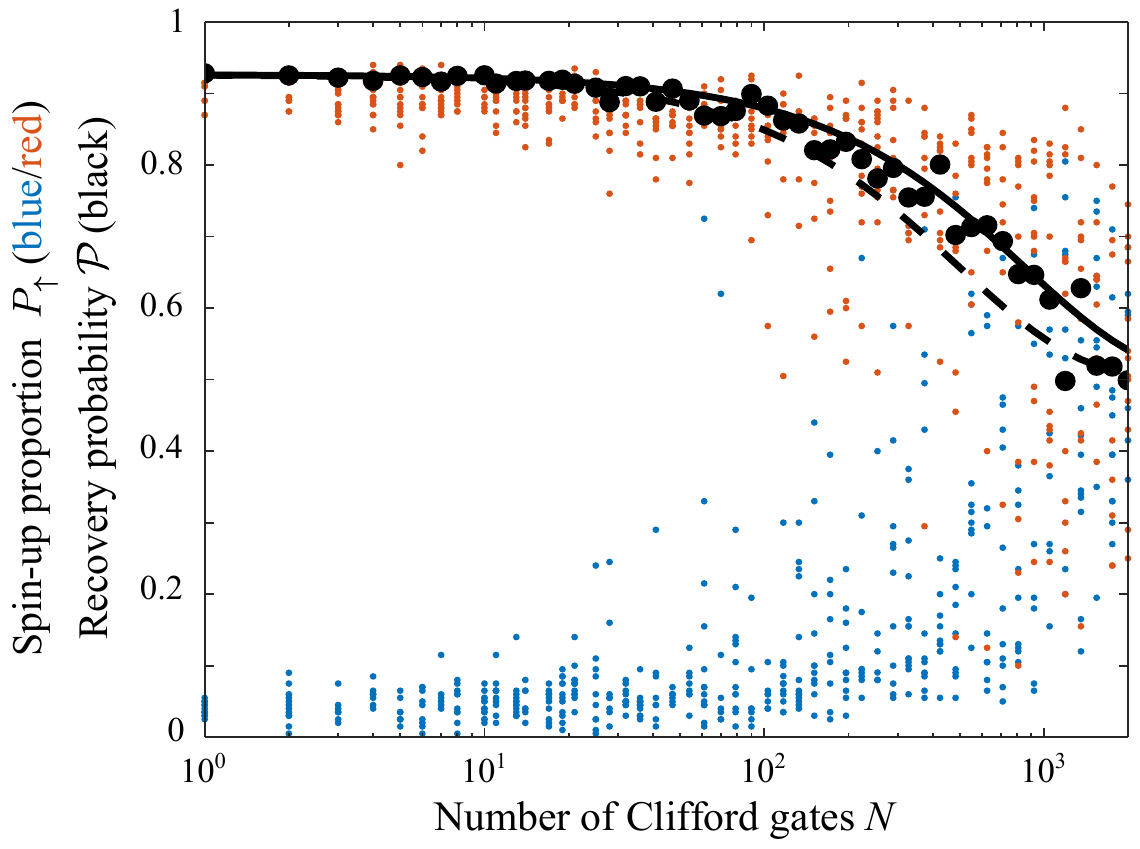}
 \caption{Randomized benchmarking with optimized pulse length calibration. Each of the small dots correspond to the $P_\uparrow$ extracted from $200$ repetitions of a sequence; here, red(blue) dots correspond to sequences where the final state was chosen to be $\ket{\uparrow}$($\ket{\downarrow}$). Large black dots correspond to the overall correct recovery probability $\mathcal{P}$ as described in the main text. The solid line is a fit to the data using \eref{eq:rb}, yielding $C_0 = 0.4265(13)$ and $p = 0.99882(16)$, corresponding to $\mathcal{F}_\mathrm{G} = 99.942(8)$. The fit is weighted with the inverse of the unbiased sample variance at each $N$. The dashed line uses $p = 0.998$, corresponding to the previously measured $\mathcal{F}_\mathrm{G} = 99.9$\%~\cite{Muhonen2015}, scaled with the same $C_0$ for comparison.\label{fig:rb}}
\end{figure}

To confirm the improvement in the gate calibration, we perform randomized benchmarking using the optimized calibration protocol. The randomized benchmarking protocol was implemented using the same Clifford gate set as in reference~\cite{Muhonen2015}. The protocol tests sequences with increasing number of Clifford gates $N$. To construct the sequences, a set of $N$ Clifford gates is selected at random; a final state ($\ket{\uparrow}$ or $\ket{\downarrow}$) is also chosen at random and a final gate is added to the random gate sequence such that the spin is flipped to this final state. This sequence is repeated $200$ times to compute $P_\uparrow$. For each $N$, $20$ different random sequences are measured. From the data sets corresponding to each $N$, we can extract the overall probability of recovering the correct state $\mathcal{P} = 0.5(\bar{P}_\uparrow^{(\uparrow)}+(1-\bar{P}_\uparrow^{(\downarrow)}))$, where $\bar{P}_\uparrow^{(\uparrow)}$ is the mean value of $P_\uparrow$ from sequences where the final state was chosen to be $\ket{\uparrow}$ ($\ket{\downarrow}$ for $\bar{P}_\uparrow^{(\downarrow)}$). $\mathcal{P}(N)$ can then fitted~\cite{Fogarty2015} to:
\begin{eqnarray}
\mathcal{P} = C_0 p^N + 0.5 \label{eq:rb}
\end{eqnarray}
where $C_0$ is a constant determined by SPAM errors and $p$ determines the gate fidelity $\mathcal{F}_\mathrm{G} = (1 + p)/2$. From the results shown in \fref{fig:rb}, we extract $\mathcal{F}_\mathrm{G} = 99.942(8)$\%, setting a new gate fidelity benchmark for the $^{31}$P electron-spin qubit.

\section{Non-Markovian noise}

The accuracy of GST relies greatly on the stability of the qubit over the timescale of the experiment. Essentially, GST assumes that the qubit is `the same qubit' when each sequence is being applied. Any slow drift in the environment will reduce GST's ability to fit the data using a Markovian model, and thereby reduce the reliability of its estimates.  While GST is able to detect and crudely quantify such non-Markovian noise (e.g. slow drift results in decreasing goodness-of-fit with increasing sequence length), it is as yet unable to assign meaningful error bars to account for this noise. An analysis of the goodness-of-fit from GST reveals that the experimental dataset violates the fitted Markovian model by up to $250$ times the standard deviation returned by the fit (see supplementary GST reports for more details). This is a strong indicator that there are high levels of non-Markovian noise present in the system.

\begin{figure}
 \includegraphics[width=\textwidth]{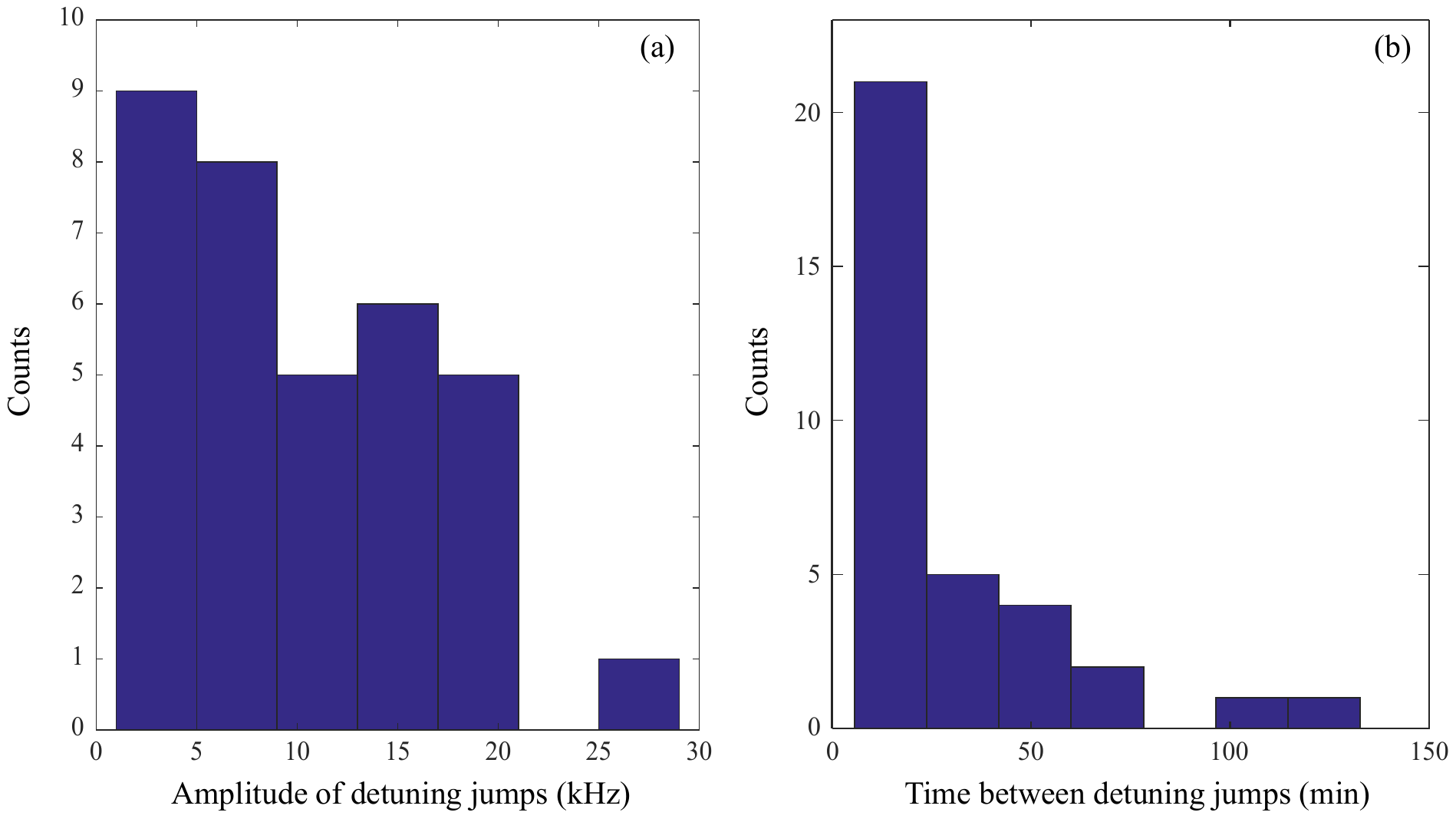}
 \caption{Statistical characterization of random jumps in the qubit resonance frequency. (a) Histogram of the amplitude of the observed frequency shifts; (b) Histogram of the time interval between frequency jumps. This data is obtained from repeated resonance frequency calibrations over a period of $\sim40$~hours. The calibration procedure is described in the main text. To obtain this dataset, a total of $791$ calibrations were performed with 3~minute intervals, and a total of $34$ frequency jumps above the the threshold were recorded. The sampling rate and total length of the Ramsey measurement is set such that the frequency resolution of the calibration is $1$~kHz and the maximum detuning detection is $100$~kHz. The mean values of each dataset are: (a) $10$~kHz and (b) 28~minutes. The Pearson correlation coefficient using the two datasets is $-0.2(3)$, which indicates little correlation between the magnitude of frequency jumps and the interval between them.\label{fig:jumps}}
\end{figure}

We attribute the majority of the non-Markovian noise to jumps on the order of $10$~kHz in the qubit resonance frequency, which happen on timescales on the order of $10$~minutes (\fref{fig:jumps}). These jumps likely arise from single nuclear spin flips from either $^{29}$Si or other ionized $^{31}$P in the vicinity of the qubit. Recalling \eref{eq:omega}, a shift in the ESR frequency will modify the Rabi oscillation frequency, which in turn will cause an error in the pulse rotation. With the $B_1$ used in our experiments, a $10$~kHz detuning will cause a $\sim0.2$\% error in pulse rotation. This is well within the accuracy capabilities of GST.

While GST and RB are expected to agree to within their respective error bars on gates with Markovian errors, they respond very differently to the slow drift that causes non-Markovian behaviour in the system. Drift in the qubit resonance frequency produces coherent (unitary) errors in the gates, but ones that vary in time. RB is largely insensitive to coherent errors of any kind \cite{Wallman2014,Sanders2015}. Large non-Markovian drifts in detuning frequency can cause the RB decay curve to become noticeably non-exponential~\cite{Veldhorst2014,Fogarty2015}; however, in the results presented here this effect is too subtle to observe. GST, on the other hand, is very sensitive to non-Markovian noise---but has no mechanism for it. GST misclassifies this kind of non-Markovian noise (caused by slow drift) as stochastic noise. Therefore, while RB underestimates the total noise, GST overestimates the stochastic noise. For this reason, simulated RB using the GST estimated gate set from the optimized system (\fref{fig:gst_results}d), predicts an average Clifford gate infidelity of $1 - \mathcal{F}_\mathrm{G} = 0.25(2)\%$. In contrast, the average Clifford gate infidelity observed in real RB experiments (\fref{fig:rb}) is $1 - \mathcal{F}_\mathrm{G} = 0.058(8)\%$. Therefore, while GST fails to correctly predict RB, this is a direct consequence of the fact that GST is able to identify non-Markovian noise (although not to model it), and correctly warns that its presence compromises the accuracy of the results. Comparison of GST and RB results indicate that non-Markovian effects currently dominate Markovian stochastic noise in the system.

Since quantum error correction schemes rely on noise being Markovian, the effects of non-Markovian noise need to be mitigated in order to use this qubit in a fault-tolerant setting. In all the experiments presented here, we monitor and calibrate the resonance frequency of the qubit by performing a Ramsey fringe experiment~\cite{Ramsey1950} to determine the detuning frequency; if the detuning frequency is found to be larger than a threshold of $5$~kHz, the output frequency of the MW signal generator is adjusted and the Ramsey fringe measurement is performed again; the process is repeated until the detuning frequency is found to be within the threshold. The calibration takes on average $\sim1$~minute to complete and is performed every $\sim20$~minutes. When performing long experiments (such as those required by GST), the experiment needs to be paused every time a resonance calibration is performed. Therefore, increasing the frequency with which the calibration is performed will unmanageably extend the total experiment duration. A different approach to minimize (but not eliminate) the impact of drift and/or non-Markovian noise is to interleave the `shots' of each GST sequence~\cite{Enk2013}. Currently, we take $100$~single-shot measurements per sequence consecutively, and run through each sequence in a single `sweep'. By performing interleaving, the measurements are taken in $100$ sequence sweeps with $1$ single-shot per sequence (or, more feasibly, repeating $100/N$ sweeps and taking $N$ shots for each sequence during each sweep). Interleaving would ensure that the data for each sequence are sampled from the full span of time for which the experiment runs. It does not eliminate non-Markovian behaviour (drift still has a significant impact on long sequences even with interleaving), but would result in a more reliable and meaningful estimate. However, this method is impractical with our current experimental setup, because the most time-consuming step in the experiment is loading a new sequence onto the arbitrary waveform generator, while repeating a measurement once a sequence is loaded is relatively much faster. Therefore, attempting to perform an adequate amount of interleaving would unmanageably increase the total duration of the experiment. Furthermore, this would not address the root of the problem: qubit drift over time that would become problematic when running real quantum circuits. Moving forward, an approach to correct this non-Markovian noise is to use dynamically corrected gates~\cite{Liu2013,Rong2014,Zhang2014}, where the gate sequence is interleaved with a dynamical decoupling sequence in order to suppress gate errors and decoherence effects from low-frequency noise sources. This approach has been successfully applied and verified to correct non-Markovian noise using GST for a trapped-ion qubit~\cite{Blume2016}, which leads us to believe that it would also be successful here. Another possible solution is to implement a Hamiltonian estimation protocol~\cite{Shulman2014}, which could potentially allow us to increase the speed and frequency of the detuning frequency calibration.

\section{Conclusion}

Gate Set Tomography is a protocol designed to characterize and optimize qubit systems. By applying GST to the $^{31}$P electron spin qubit in $^{28}$Si, we were able to identify a $4.4$\% rotation error in some of the gates. We improved the calibration method to fix this error, which in turn improved the average gate fidelity of the qubit from $99.90(2)\%$ to $99.942(8)\%$, measured via randomized benchmarking. Non-Markovian noise, originating from small jumps in the resonance frequency of the qubit, are detected by GST, and limit the performance of the qubit. The use of dynamically corrected gates should suppress the effects of non-Markovian noise, and should be first priority for future measurements. This work demonstrates that GST is capable of characterizing qubit gates to levels not previously accessible through any other experimental protocol. We envision that GST will become an increasingly important tool for validation and verification of quantum information hardware and protocols, as the community moves towards increasingly complex and high-fidelity gate operations.

\ack
We thank F. Hudson for assistance in device fabrication and design, K. Itoh for supplying the $^{28}$Si epilayer, and J. McCallum and D. Jamieson for the $^{31}$P donor implantation. This research was funded by the Australian Research Council Centre of Excellence for Quantum Computation and Communication Technology (project no. CE110001027) and the US Army Research Office (W911NF-13-1-0024). The authors acknowledge support from the Australian National Fabrication Facility. Sandia National Laboratories is a multi-program laboratory managed and operated by Sandia Corporation, a wholly owned subsidiary of Lockheed Martin Corporation, for the U.S. Department of Energy’s National Nuclear Security Administration under contract DE-AC04-94AL85000. JKG gratefully acknowledges support from the Sandia National Laboratories Truman Fellowship Program, which is funded by the Laboratory Directed Research and Development (LDRD) program.

\section*{References}
\providecommand{\newblock}{}

\end{document}